\documentclass[slac_one]{revtex4}
\usepackage{graphicx}
\usepackage{fancyhdr}
\begin{document}

\title{Prospects for Electroweak Measurements at the LHC} 

%

\author{Martin W. Gr\"unewald (on behalf of the ATLAS and CMS collaborations)}
\affiliation{University College Dublin and University of Ghent}

\begin{abstract}

The prospects for electroweak measurements at the Large Hadron
Collider (LHC) are discussed.  In addition to high-luminosity results,
special emphasis is placed on early start-up measurements with a total
luminosity ranging from 10/pb to 100/pb, using the general-purpose
detectors ATLAS and CMS and their initially larger calibration and
alignment uncertainties.  Topics discussed here include inclusive W
and Z production, W-boson mass, Z forward-backward asymmetry,
Z-plus-jets production and di-boson production, the latter
constraining trilinear electroweak gauge couplings.
(Invited talk at the 34th ICHEP, Philadelphia, USA, July/August 2008)

\end{abstract}

\maketitle

\thispagestyle{fancy}


\section{INTRODUCTION}

After more than a decade of construction, the LHC has successfully
been turned on in September of 2008.  Following the end-of-year
shutdown, 2009 will see first proton-proton collisions, at
centre-of-mass energies of perhaps 900~GeV (injection), 10~TeV and
14~TeV, with a total luminosity reaching 1/fb or more.  This paper
summarises the prospects for electroweak measurements at the LHC with
the two general-purpose detectors ATLAS and CMS.

Both ATLAS and CMS are built for high-pt physics at the LHC.  Their
most visible difference is in their size and magnetic field
arrangement: while ATLAS is dominated by a 2.5-3.5 Tesla toroid
combined with a central 2 Tesla solenoid, CMS uses a large
superconducting solenoid of 4 Tesla, with the outer return flux
instrumented with muon chambers. The muon systems of the detectors
provide a coverage of 2.7 (ATLAS) and 2.4 (CMS) in pseudo-rapidity
$|\eta|$. The momentum resolution for muons of 1~TeV transverse
momentum (pt) is 8\% in ATLAS and 5\% in CMS. The central tracking
system of both detectors covers $|\eta|<2.5$, providing resolutions of
3.7\% (ATLAS) and 1.5\% (CMS) for tracks with pt of 100 GeV. The
electromagnetic crystal calorimeter of CMS provides a superior energy
resolution for electromagnetic showers, $3\%$/$\sqrt{E}\oplus0.25\%$
within $|\eta|<3.0$, versus $10\%/\sqrt{E}\oplus0.5\%$ within
$|\eta|<3.2$ for the ATLAS calorimeter.  Conversely, the ATLAS
hadronic calorimeter ($|\eta|<4.9$) provides for a better energy
resolution compared to CMS ($|\eta|<5.2$). For jets and combining the
electromagnetic and hadronic calorimeter response, energy resolutions
of $60\%$/$\sqrt{E}\oplus3\%$ (ATLAS) and $70\%$/$\sqrt{E}\oplus8\%$
(CMS) are expected. The resolutions vary strongly with $\eta$; more
details can be found in~\cite{DETECTORS}.

Compared to the Tevatron centre-of-mass energy of 2 TeV, the factor 7
higher energy of the LHC results in cross sections for the production
of heavy particles to increase by up to one or two orders of
magnitude.  With an instantaneous luminosity of $10^{33}/cm^2/sec$,
the typical LHC event rates are 150 Hz W, 50 Hz Z and 1 Hz $t \bar
t$. In an initial luminosity of 10/pb, we expect 150000 $W\to e\nu$
events, 15000 $Z\to ee$ events and 10000 $t \bar t$ events.  Hence
very soon after LHC start-up, event statistics of electroweak final
states will not limit the measurements.

The signature of $W\to\ell\nu$ events is given by a high-energy
lepton, missing (transverse) energy due to the neutrino, and a
hadronic system recoiling against the decaying W boson. Inclusive
$Z\to\ell^+\ell^-$ events contain a pair of high-energy leptons of
opposite electric charge, no missing energy but again a hadronic
recoil system. In both cases, the hadronic recoil spans the region
from being very soft to hard, possibly even leading to one or more
jets.

\section{INCLUSIVE W/Z PRODUCTION}

Inclusive W/Z production is considered as a fundamental benchmark
process, also at the LHC, where it will be measured in the new regime
of 10~TeV and 14~TeV center-of-mass energy. The cross section at
14~TeV is about 180nb and 60nb for W and Z production, respectively,
and is theoretically one of the best understood cross sections at
hadron colliders, especially concerning uncertainties due to radiative
corrections and parton distribution functions (PDFs). The W/Z process
has the potential to become a high-pt reaction for the determination
of the luminosity at the few \% uncertainty level.  Further, inclusive
W/Z production serves as the starting point for more detailed
analyses, such as measuring the boson pt spectrum, looking for
additional jets, or measurements of the Z-decay asymmetry and W-boson
mass and width.  In particular, Z events provide a crucial calibration
source, given the precise knowledge of the Z mass and width as
measured at LEP.

As an example, the ATLAS lepton identification and event selections
designed for the LHC start-up, for 50/pb of luminosity, are listed
here~\cite{ATLAS-EWK}: electrons are identified within $|\eta|<2.4$
and muons within $|\eta|<2.5$. Quality criteria on electron candidates
are assigned, such as ``loose'' (using the calorimetric showershape),
``medium'' (adding track and track matching requirements), and
``tight'' (sharpening these requirements). For muon candidates,
isolation, the amount of activity in a cone around the muon, is used.

The selection of $W\to\ell\nu$ events requires either a medium-quality
electron or an isolated muon. The transverse energy of the lepton and
the missing transverse energy must both be larger than 25 GeV, and the
transverse mass of the lepton-missing energy system must be larger
than 40 GeV.  In $Z\to\ell^+\ell^-$, two charged leptons are present
so that the lepton identification is relaxed: either two loose
isolated electrons or two isolated muons of opposite charge. In case
of electrons, the transverse energy must be larger than 15 GeV and the
invariant di-electron mass must be in the range from 80 to 100 GeV.
In case of muons, the transverse momentum must be larger than 20 GeV,
and the di-muon mass be in a $\pm$20~GeV window around the Z mass
value.  The selection and trigger efficiencies range from 60\% to
90\%. Both experiments exploit data-driven determinations using
tag-and-probe on Z decays.

The expected statistical and systematic uncertainties on the event
numbers (rate) are as follows, using ATLAS 50/pb, ATLAS 1/fb and CMS
1/fb selections~\cite{ATLAS-EWK,CMS-Zee,CMS-Zmm}: For W events, the
statistical uncertainties are 0.2\%, 0.04\% and 0.04\%, while the
systematic uncertainties are projected as 3.1-5.2\%, 2.4\% and 3.3\%.
The systematic uncertainty is dominated by the missing energy
determination.  For the Z rate, the corresponding numbers are 0.8\%,
0.2\% and 0.13\% statistical uncertainty, and 3.2-3.6\%, 1.3\% and
2.3\% systematic uncertainty. The theoretical systematic uncertainty
is dominated by PDFs and the underlying boson pt distribution.  Thus
even with a small amount of luminosity at LHC start-up, the rate
measurements are dominated by systematic uncertainties. The systematic
uncertainties will decrease with improved understanding of the
detectors, but slower than the statistical uncertainty.

In order to turn the rate into a cross section, a luminosity
determination is needed, typically obtained by measuring forward
scattering. The uncertainty on the luminosity from this method is
estimated to be 10\% initially, decreasing to about 5\% in the long
term. It is thus attractive to use W/Z production as an alternative
luminosity reaction, because a smaller uncertainty can be achieved.
Further, using a high-pt process similar to other signal processes,
e.g., $t\bar t$ production, theoretical uncertainties due to PDFs and
other issues partially cancel in the ratio.

\subsection{W Mass}

The mass of the W boson is a fundamental parameter of the electroweak
Standard Model; in particular, together with the mass of the top
quark, it constrains the mass of the as yet undiscovered Higgs
boson~\cite{PeterRenton}.  The W-boson mass and width is measured
precisely at LEP-2~\cite{PeterRenton} and by the Tevatron experiments
CDF and D\O~\cite{AshutoshKotwal}.  The measurement requires a clean
sample of W decays, thus tighter quality criteria on the lepton
identification are imposed. In case of ATLAS~\cite{ATLAS-WMass}, one
requires for $W\to e\nu$ exactly one isolated tight electron
candidate, and for $W\to\mu\nu$ one isolated muon candidate. The
transverse energy of the lepton and the missing transverse energy must
both exceed 20 GeV.

Already in 15/pb of luminosity, 67000 $W\to e\nu$ and 120000
$W\to\mu\nu$ events will be selected, together with 3000 $Z\to ee$ and
10000 $Z\to\mu\mu$ events.  The W mass will be extracted from the
Jacobian peak observed in the transverse mass of the lepton-neutrino
system, or the transverse energy of the charged lepton. The Z events
are a crucial source of calibration for the lepton energy scale (known
Z mass) and energy resolution (known Z width), and used as well in the
determination of the differential lepton reconstruction efficiency.
The low-luminosity ATLAS study shows how well the energy scale and
resolution can be monitored through Z events as a function of pseudo
rapidity, for example the required corrections due to transition
effects between central and endcap calorimeters. For 15/pb of
luminosity, a W-mass uncertainty ranging from 160 to 240 MeV is
expected. While this is not meant to be competitive with current
measurements at LEP-2 and the Tevatron, it serves to establish the
W-mass analysis at the LHC.

Requiring higher luminosities, novel techniques were studied by the
CMS collaboration to measure the W-boson mass through templates
generated from data (Z events), thus no longer relying on MC
simulations~\cite{CMS-TDR}. Two possibilities are
studied~\cite{Giele}: (i) an event-by-event transformation to change a
Z event into a W event corresponding to a trial value of the W boson
mass: one takes a $Z\to\ell\ell$ event, boosts it to the Z
rest-system, rescales the lepton momenta by the ratio
$M_W(trial)/M_Z(LEP)$, removes one lepton to mimic a neutrino, and
boosts back to the detector system; and (ii) transformation of
distributions, such as the lepton pt distribution~\cite{Giele}.  The
advantage of these methods lies in the fact that Z events from data
rather than MC simulations are used, so that many systematic errors
disappear and only the residual W-Z differences need to be
studied. These methods require high luminosity as Z events from data
are used. For 1/fb and 10/fb of luminosity, CMS expects statistical
uncertainties of 40 and 15 MeV, with experimental systematic errors of
40 and 20 MeV, and PDF uncertainties of 20 and 10 MeV, respectively.

\subsection{Z Forward-Backward Asymmetry}

Even in Z production in proton-proton collisions, a forward-backward
asymmetry of the Z decay products is expected. The Z is formed by a
quark-antiquark pair; while the anti-quark always arises from the sea,
the quark may also be a valence quark which on average carries a
higher momentum than sea quarks. Thus the boost direction indicates
the quark direction at high rapidities. In a sample of high-rapidity
electron pairs, an asymmetry is observed which can be interpreted as a
measurement of the effective electroweak mixing angle,
$\sin^2\theta_{eff}$, similar to the asymmetries measured at LEP and
at the Tevatron.  With 100/fb of luminosity, ATLAS~\cite{ATLAS-AFB}
expects a measurements of $\sin^2\theta_{eff}$ with a statistical
precision of 0.00015 and a systematic uncertainty of 0.00024,
comparable to the uncertainty of the world average dominated by LEP
and SLD~\cite{PeterRenton}.  The D\O\ experiment at the Tevatron has
made a measurement with a statistical precision of 0.0018 and
systematic uncertainty of 0.0006 using 1.1/fb~\cite{SusanBlessing}.
The systematic uncertainties are by far dominated by PDF-related
uncertainties, but the knowledge of PDFs is expected to improve
through measurements at the Tevatron, HERA, and also LHC (e.g., W
asymmetry measurements).

\section{Z PLUS JETS}

Z production accompanied by jets serves as a test of perturbative QCD
but is also a major background in searches for new physics, thus a
good understanding of this process is required.  The
ATLAS~\cite{ATLAS-ZJets} selection designed for a luminosity of 1/fb
uses the standard $Z\to ee$ selection, while jets are clustered in a
cone of $R=0.4$ and considered within the fiducial volume of
$|\eta|<3$.  Jets are required to have a pt larger than 40 GeV. The
lepton-jet separation must exceed $R=0.4$.  The background from
heavy-particle final states ($Z\to\tau\tau$, $W$, $t\bar t$) is taken
from MC, while the QCD multi-jet background is derived from data,
where MC simulations indicate that the expected multi-jet background
fraction is independent of the jet pt. With 1/fb of luminosity, up to
4 jets can be observed in rate and pt spectrum, allowing to test MC
models.  A CMS study~\cite{CMS-Zbb} specifically investigates $Z+b\bar
b$ production resulting in a signature of two leptons and two b-jets.
The background consists of Drell-Yan production plus light jets,
$Z+c\bar c$ and $t \bar t$ production.  The selection requires two
isolated leptons of opposite charge and pt larger than 20 GeV, and at
least two b-tagged jets within $|\eta|<2.4$ and pt larger than 30~GeV.
In order to reject $t\bar t$ events the missing energy must be smaller
than 50 GeV.  Within 100/pb, this results in a cross section
determination with a statistical (systematic) uncertainty of 15\%
(23\%), the latter dominated by the jet energy scale and missing
energy systematic.

\vskip -2mm

\section{DI-BOSON PRODUCTION}

\vskip -2mm

Pair-production of electroweak gauge bosons tests the triple gauge
boson couplings of the electroweak Standard Model. Within the SM, the
trilinear vertices $WW\gamma$ and $WWZ$ occur, while those involving
only neutral gauge bosons, $\gamma$ and Z, are absent.  The charged
triple gauge couplings (TGCs) are usually taken as $g^V_1$, $\kappa^V$
and $\lambda^V$ for $V=\gamma,Z$; they are related, for $V=\gamma$, to
the magnetic dipole and electric quadrupole moment of the W
boson. Within the SM, their values are $g_1=\kappa=1$ and $\lambda=0$.
Di-boson production leads to final states containing charged leptons
from W/Z decay and phtons.  The photon identification is similar to
the electron identification except for a veto on charged tracks
matching the calorimetric cluster of the photon candidate.

CMS is using a cut-based analyses~\cite{CMS-TDR} while ATLAS studied
in addition a boosted decision tree with improved sensitivity compared
to their cut-based analysis~\cite{ATLAS-DiBoson}. The number of
selected events for signal (background) obtained using the ATLAS
boosted-decision-tree selections on a total luminosity of 1/fb are:
$W\gamma$: 3770 (2525) and $Z\gamma$ 1118 (616), with the background
dominated by W/Z plus fake photons; $WW$: 469 (92) yielding a signal
significance of 10 standard deviations already with 0.1/fb of
luminosity and a 20\% background uncertainty; $WZ$: 128 (16) yielding
a 5.8 sigma significance for 0.1/fb and a 20\% background uncertainty;
$ZZ\to 4\ell$ 13.3 (0.2) for a signal significance of 6.8 sigma in
1/fb, and $ZZ\to 2\ell 2\nu$ 10.2 ($5.2\pm2.6$).  Anomalous TGCs lead
to increased cross sections especially at high boson pt and di-boson
transverse mass, allowing to set limits.  Systematic uncertainties in
95\% CL limits on anomalous TGCs become relevant only for luminosities
of 30/fb or higher.

\section{CONCLUSION}

The LHC will provide proton-proton collisions in 2009.  The four
detectors ATLAS, CMS, LHCb and ALICE are eagerly awaiting collision
data.  Both luminosity and cross sections at the LHC are much higher
compared to earlier experiments, hence there will be no lack of
statistics and sensitivity to rare processes such as ZZ production.
The prospects of electroweak measurements are exciting due to
high-performance detectors, allowing to place tight constraints on the
electroweak Standard Model, through measurements of production rates,
masses and couplings of the electroweak gauge bosons.  To exploit the
data it is important to understand the early data and detectors
quickly.

\subsection*{Acknowledgments} \label{Ack}

It is a pleasure to thank my colleagues from the CMS and ATLAS
collaborations, notably Juan Alcaraz and Tom LeCompte, for discussing
results and answering patiently my questions.

$ $ \vskip -2cm $ $

\end{document}